# Inter-Cloud Data Security Strategies

Sugata Sanyal
**Corporate Technology Office**
**Tata Consultancy Services**
Mumbai, INDIA
sugata.sanyal@tcs.com

Parthasarathy P. Iyer (Corresponding Author)
**Self-Inventions-R&D**
Mumbai, INDIA
iyerparth@rediffmail.com

*Abstract- Cloud computing is a complex infrastructure of software, hardware, processing, and storage that is available as a service. Cloud computing offers immediate access to large numbers of the world's most sophisticated supercomputers and their corresponding processing power, interconnected at various locations around the world, proffering speed in the tens of trillions of computations per second. Information in databases and software scattered around the Internet. There are many service providers in the internet, we can call each service as a cloud, each cloud service will exchange data with other cloud, so when the data is exchanged between the clouds, there exist the problem of security. Security is an important issue for cloud computing, both in terms of legal compliance and user trust, and needs to be considered at every phase of design. In contrast to traditional solutions, where the IT services are under proper physical, logical and personnel controls, Cloud Computing moves the application software and databases to the large data centers, where the management of the data and services may not be trustworthy. This unique attribute, however, poses many new security challenges. Cloud computing seems to offer some incredible benefits for communicators.*

## I. INTRODUCTION

The availability of an incredible array of software applications, access to lightning-quick processing power, unlimited storage, and the ability to easily share and process information. All of this is available through your browser any time you can access the Internet. While this might all appear enticing, there remain issues of reliability, portability, privacy, and security. The brave new world of Cloud computing offers many benefits provided that the privacy and security risks are recognized and effectively minimized[1][2].

1) Limitless flexibility: With access to millions of different pieces of software and databases, and the ability to combine them into customized services, users are better able to find the answers they need, share their ideas, and enjoy online games, video, and virtual worlds.

2) Better reliability and security: Users no longer have to worry about their hard drives crashing or their laptops being stolen;

3) Enhanced collaboration: By enabling online sharing of information and applications, the Cloud offers users new ways of working and playing together. [3]

4) Portability: Users can access their data and tools wherever they can connect to the Internet;

5) Simpler devices: With data and the software being stored in the Cloud, users do not need a powerful computer. They can interface using a cell phone, a PDA, a personal video recorder, an online game console, their cars, or even sensors built into their clothing.

We can only have the full benefits of Cloud computing if we can address the very real privacy and security concerns that come along with storing sensitive personal information in databases and software scattered around the Internet[4][5].

### A. Cloud Types

*1. Hybrid clouds:*
Hybrid clouds are a combination of the public and the private cloud using services that are in both the public and private space. Management responsibilities are divided between the public cloud provider and the business itself.

Using a hybrid cloud, organizations can determine the objectives and requirements of the services to be created and obtain them based on the most suitable alternative [6].

*2. Public clouds:*

Public clouds are available to the general public or a large industry group and are owned and provisioned by an organization selling cloud services. A public cloud is what is thought of as the cloud in the usual sense; that is, resources dynamically provisioned over the Internet using web applications from an off-site third-party provider that supplies shared resources and bills on a utility computing basis.

*3. Private clouds:*

Private clouds exist within your company's firewall and are managed by your organization. They are cloud services you create and control within your enterprise. Private clouds offer many of the same benefits as the public clouds — the major distinction being that your organization is in charge of setting up and maintaining the cloud.

*4. Community cloud:*

Community cloud: The cloud infrastructure is shared by several organizations and supports a specific community that has shared concerns (e.g., mission, security requirements, policy, and compliance considerations). It may be managed by the organizations or a third party and may exist on premise or off premise [7].

*B. Cloud Computing Models*

While there is a significant buzz around Cloud Computing, there is little clarity over which offerings qualify or their interrelation. The key to resolving this confusion is the realization that the various offerings fall into different levels of abstraction [8] [9], aimed at different market segments.

1) Infrastructure-as-a-Service (IaaS): At the most basic level of Cloud Computing offerings. These instances essentially behave like dedicated servers that are controlled by the developers. So, once a machine reaches its performance limits, the developers have to manually instantiate another machine and scale their application out to it. This service is intended for developers who can write arbitrary software on top of the infrastructure with only small compromises in their development methodology.

2) Platform-as-a-Service (PaaS): One level of abstraction above, services like Google App Engine provides a programming environment that abstracts machine instances and other technical details from developers. The programs are executed over data centers, not concerning the developers with matters of allocation. In exchange for this, the developers have to handle some constraints that the environment imposes on their application design.

3) Software-as-a-Service (SaaS): At the consumer-facing level are the most popular examples of Cloud Computing, with well-defined applications offering users online resources and storage. This differentiates SaaS from traditional websites or web applications which do not interface with user information (e.g. documents) or do so in a limited manner. Popular examples include Microsoft's (Windows Live) Hotmail, office suites such as Google Docs.

| SAAS | PAAs | IAAS |
|---|---|---|
| Zoho, Salesforce.com, Google Apps | Windows Azure, Google App Engine, Aptana Cloud | Drop box, Amazon Web Services, Mozy, Akamai |

Table I. Cloud Providers with Cloud Service

## II. Models

*A. Security Management Techniques*

**MAC-**In the Mandatory Access Control (MAC) mode, users are given permissions to resources by an administrator. Only an administrator can grant permissions or right to objects and resources. Access to resources is based on an object's security level, while users are granted security clearance. Only administrators can modify an object's security label or a user's security Clearance [10].

**DAC-**In the Discretionary Access Control (DAC) model, access to resources is based on user's identity. A user is granted permissions to a resource by being placed on an access control list (ACL) associated with resource. An entry on a resource's ACL is known as an Access Control Entry (ACE). When a user (or group) is the owner of an object in the DAC model, the user can grant permission to other users and groups [11]. The DAC model is based on resource ownership. Introduced by Ferrailo and Kuhn, RBAC is used for advanced access control. It reduces the complexity and cost of security administration in large networked applications role-based access control (RBAC) is an approach to restricting system access to authorized users[12][13][14]. It is a newer alternative approach to mandatory access control (MAC) and discretionary access control (DAC). RBAC is sometimes referred to as role-based security.

RBAC defines three models:
1. Core defines the basic elements listed below
2. Hierarchical defines the relation between roles
3. Constraint defines condition in role assignment:
Static separation of duties defines mutually disjoint user agreements. Dynamic separation of duties limits permissions to user sessions. A subject can have multiple roles. A role can have multiple subjects. A role can have much permission. Permission can be assigned too many roles. Three primary rules are defined for RBAC:
1. Role assignment: A subject can execute a transaction only if the subject has selected or been assigned a role.
2. Role authorization: A subject's active role must be authorized for the subject. With rule 1 above, this rule ensures that users can take on only roles for which they are authorized.
3. Transaction authorization: A subject can execute a transaction only if the transaction is authorized for the subject's active role.
With rules 1 and 2, this rule ensures that users can execute only transactions for which they are authorized. Improved RBAC: If the requesting cost of service is less than budget limit then role is not changed, but if it is greater than or equal to the budget then role is changed to role for higher version performance [15]. It can decrease unnecessary purchases of computer resources for service upgrading.

### *B. Security Concerns Of Cloud Computing*
While cost and ease of use are two great benefits of cloud computing, there are significant security concerns that need to be addressed when considering moving critical applications and sensitive data to public and shared cloud environments [16]. To address these concerns, the cloud provider must develop sufficient controls to provide the same or a greater level of security than the organization would have if the cloud were not used.

### *C. Cloud Computing Is Based On Five Attributes*

1. Multi-tenancy (shared resources): Cloud computing is based on a business model in which resources are
shared (i.e., multiple users use the same resource) at the network level, host level, and application level.

2. Massive scalability: Cloud computing provides the ability to scale to tens of thousands of systems, as well as the ability to massively scale bandwidth and storage space.

3. Elasticity: Users can rapidly increase and decrease their computing resources as needed.

4. Pay as you used: Users pay for only the resources they actually use and for only the time they require them.

5. Self-provisioning of resources: Users self-provision resources, such as additional systems (processing capability, software, storage) and network resources.

Cloud computing eliminates the costs and the complexity of buying, configuring, and managing the hardware and software needed to build and deploy applications; these applications are delivered as a service over the Internet[17][18].Cloud services exhibit five essential characteristics that demonstrate their relation to, and differences from, traditional computing approaches such as (1) On-demand self-service, (2) Broad network access, (3) Resource pooling, (4) Rapid elasticity, and (5) Measured service [19].

Cloud computing often leverages Massive scale, Homogeneity, Virtualization, Resilient computing (no stop computing), Low cost/free software, Geographic distribution, Service orientation Software and Advanced security technologies. Cloud computing combines a number of computing concepts and technologies for Service Oriented Architecture (SOA), which may include Web 2.0 and the virtualization of services and communication infrastructure[20][21]. These technologies have allowed cloud customer organizations to achieve improved utilization and efficiency of their service providers" infrastructure through the controlled sharing of computing resources with other customers (multi-tenancy); and, greater flexibility to scale up and down IT services. In some respects, cloud computing represents the maturing of these technologies and is a marketing term to represent that maturity and the cloud services they provide.

## CONCLUSION

In this paper we have assessed some of the key issues involved, and set out the basis of some approaches that we believe will be a step forward in addressing security issues. Cloud providers need to safeguard the privacy and security of personal data that they hold on behalf of organizations and users. Security is a major obstacle in the progress of cloud computing. Numerous solutions have been proposed to overcome it. Some solutions focus on providing protection to the cloud system where the data has been stored. Others try to track possible attacks on the data. Cloud providers often have several powerful servers and resources in order to provide appropriate services for their users but cloud is at risk similar to other Internet-based technology. Various recovery techniques have also been proposed. The system gives user the chance to check his/her own data while allowing the cloud service provider to maintain the transparency of its infrastructure.

Cloud computing may raise privacy and security concerns, but this growing practice of offloading computation and storage to remote data centers run by companies such as Google, Microsoft, and Yahoo, could have one clear advantage of better energy efficiency. The system also guarantees data recovery in case of corruption of data, or in case of failure of entire server. This system can be interfaced with any cloud system with minimum modifications in the present cloud system, thus making it cost effective with respect to infrastructure. This gives the user back his/her authority over his own data.